\documentclass[prb,twocolumn,amsmath,amssymb,superscriptaddress,longbibliography]{revtex4-2}
\usepackage[
    breaklinks,
    colorlinks,
    bookmarks=false,
    citecolor=blue,
    linkcolor=red,
    urlcolor=blue
    ]{hyperref}
\usepackage{blindtext,enumitem,graphicx,dcolumn,color,xcolor,amsmath,braket,bm,changes,chemformula,cleveref}
\usepackage{tikz-feynman}
\usepackage[english]{babel}

\graphicspath{{figures_prl/}}
\usepackage{blindtext}
\usepackage{changes}

\usepackage{tabularray}

\hypersetup{
    colorlinks,
    allcolors=blue, 
}

\usepackage{orcidlink} 

\crefname{equation}{Eq.}{Eqs.}
\Crefname{equation}{Eq.}{Eqs.}
\crefname{figure}{Fig.}{Figs.}
\Crefname{figure}{Fig.}{Figs.}
\crefname{section}{Sec.}{Secs.}
\Crefname{section}{Sec.}{Secs.}

\newcommand{\canted}{c-120$^{\circ}$}

\newcommand{\JtJhVBSCol}{purple}
\newcommand{\cantCol}{orange}
\newcommand{\plaquetteCol}{blue}
\newcommand{\JdCol}{green}
\newcommand{\neelCol}{red}

\makeatletter
\appto{\appendix}{%
  \@ifstar{\def\theequation@prefix{A.}}%
          {}%
}
\makeatother

\begin{document}

\makeatletter
\renewcommand{\fnum@figure}{FIG~\thefigure}
\renewcommand{\fnum@table}{TAB.~\thetable}
\makeatother

\title{Competing Paramagnetic Phases in the Maple-Leaf Heisenberg Antiferromagnet}

\author{Paul L. Ebert\,\orcidlink{0000-0003-1614-6920}}
\email{pebert@pks.mpg.de} 
\affiliation{Max Planck Institute for the Physics of Complex Systems, N\"othnitzer Strasse 38, Dresden 01187, Germany}

\author{Yasir Iqbal\,\orcidlink{https://orcid.org/0000-0002-3387-0120}}
\affiliation{Department of Physics and Quantum Centre of Excellence for Diamond and Emergent Materials (QuCenDiEM), Indian Institute of Technology Madras, Chennai 600036, India}

\author{Alexander Wietek\orcidlink{0000-0002-4367-3438}}
\email{awietek@pks.mpg.de}
\affiliation{Max Planck Institute for the Physics of Complex Systems, N\"othnitzer Strasse 38, Dresden 01187, Germany}
\affiliation{Department of Physics and Quantum Centre of Excellence for Diamond and Emergent Materials (QuCenDiEM), Indian Institute of Technology Madras, Chennai 600036, India}

\begin{abstract}
    We establish a remarkably rich ground state phase diagram in the maple-leaf lattice spin-$1/2$ Heisenberg antiferromagnet as a function of the three symmetry-inequivalent nearest-neighbor bonds using exact diagonalization and tower-of-states analysis on clusters up to $N=36$ sites.
    Besides a hexagonal plaquette state, a star-shaped valence bond solid state is discovered in close vicinity to the (canted) $120^\circ$ magnetic phase, strongly reminiscent of a de-confined critical point or Dirac spin liquid scenario on the triangular lattice antiferromagnets.
    Moreover, an exact dimer product-state is observed next to a collinear N\'eel-state, similar to the Shastry-Sutherland model.
    All identified phases compete in a parameter regime close to the isotropic point, providing a promising region for spin liquids to emerge.
    By analyzing Gutzwiller-projected wave-functions we identify a sliver of parameter regime where a gapped $\mathbb{Z}_{2}$ spin liquid {\it Ansatz} is in astonishing agreement with the exact $N=36$ ground state.
    This rich competition of paramagnetic phases demonstrates that the maple-leaf antiferromagnet is a promising platform for exotic states of matter and quantum critical phenomena. 

\end{abstract}

\date{\today}

\maketitle

\paragraph*{Introduction.} 

    The ``maple-leaf lattice'' (MLL) refers to a 1/7 site-depleted triangular lattice~\cite{betts:1995} at a coordination number of $z=5$.
    This renders the MLL a prototypical platform between the triangular ($z=6$) and kagome ($z=4$) lattices, the (extended) Heisenberg models on both of which are known to host intriguing phases such as different types of quantum spin liquids (QSLs)~\cite{jiang:2023, shijie:2019, wietek:2017, yang:2024, wietek:2015, sun:2024}.
    In view of this, a novel aspect of the MLL is the absence of reflection symmetries, which are normally an integral part of QSL classification~\cite{bieri:2016}.
    In analogy to the Shastry-Sutherland lattice, the Heisenberg antiferromagnet (HAF) on the MLL hosts an exact dimer eigenstate throughout its phase diagram~\cite{ghosh:2022} as well as N\'eel-ordered ground states.
    The MLL structure occurs in various minerals~\cite{hawthorne:1993, fennell:2011, mills:2014, makuta:2021, kampf:2013, haraguchi:2021} and synthetic compounds~\cite{cave:2006, Aguilar-Maldonado:2025, aliev:2012, haraguchi:2018, venkatesh:2020, saha:2023}.
    
\begin{figure}
        \centering
        \includegraphics[width=0.9\linewidth]{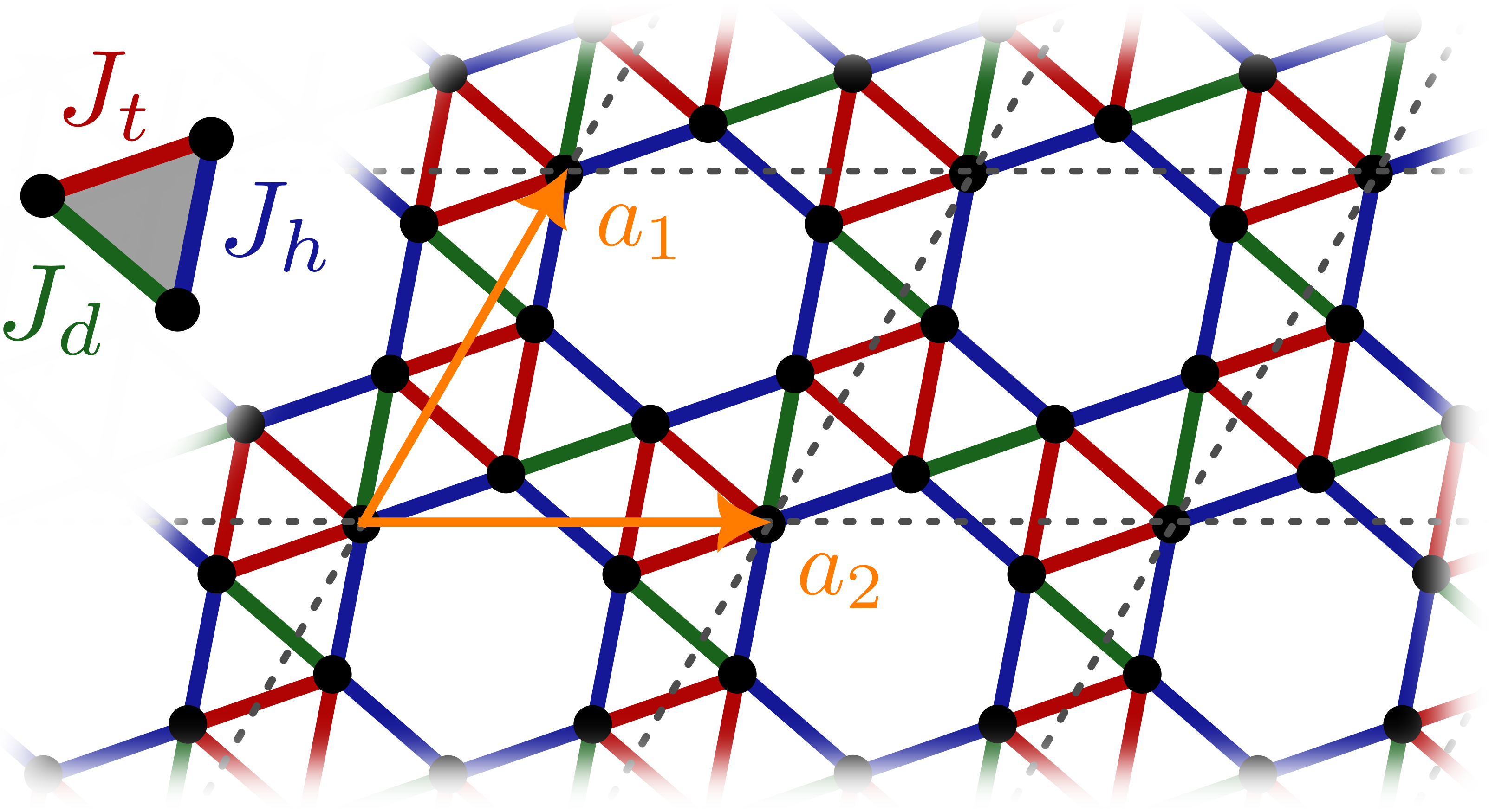}
        \caption{\label{fig:lattice}
            Maple-leaf lattice with three NN bonds: $J_t$ (red triangles), $J_h$ (blue hexagons), $J_d$ (green ``dimers''); possible six-atom units cells are highlighted.
        }
    \end{figure}

    We consider the nearest-neighbor (NN) Heisenberg model on the MLL with three symmetry-inequivalent bonds: $J_t$ (red triangles), $J_h$ (blue hexagons) and $J_d$ (green ``dimers''), as illustrated in Fig.~\ref{fig:lattice}.
    The $T=0$ phase diagram of the $0 \leq J_t, J_h, J_d$ HAF has only been studied along the $J = J_t= J_h$ line where $J_d/J$ is varied from the $J_d = 0$ ruby lattice to the $J_d = \infty$ isolated dimer limit, crossing the isotropic point with all equal couplings at $J_d / J = 1$.
    Strong efforts by various numerical techniques~\cite{schulenburg:2000, schmalfuss:2002, farnell:2011, farnell:2014, farnell:2018, gresista:2023, beck:2024, gembe:2024, schmoll:2025, schaefer:2025, richter:2004, nyckees:2025} established that the exact $J_d$-dimer eigenstate becomes the ground state above $1.45 \lesssim J_d/J$~\cite{farnell:2011, gresista:2023, beck:2024, schmoll:2025, nyckees:2025}, while the nature of the phase(s) between $0 \leq J_d/J \leq 1.45$ remains elusive, including the isotropic point.
    Proposals range from a ``canted 120$^{\circ}$'' (\canted{}) order~\cite{schmalfuss:2002, schulenburg:2000, farnell:2011, gresista:2023, beck:2024, nyckees:2025}, reminiscent of the classical ground state~\cite{farnell:2011, schmalfuss:2002, schulenburg:2000, ghosh:2025b}, to paramagnetism~\cite{gresista:2023, schmoll:2025, schaefer:2025}.

    \begin{figure*}[!ht]
        \centering
        \includegraphics[width=\textwidth]{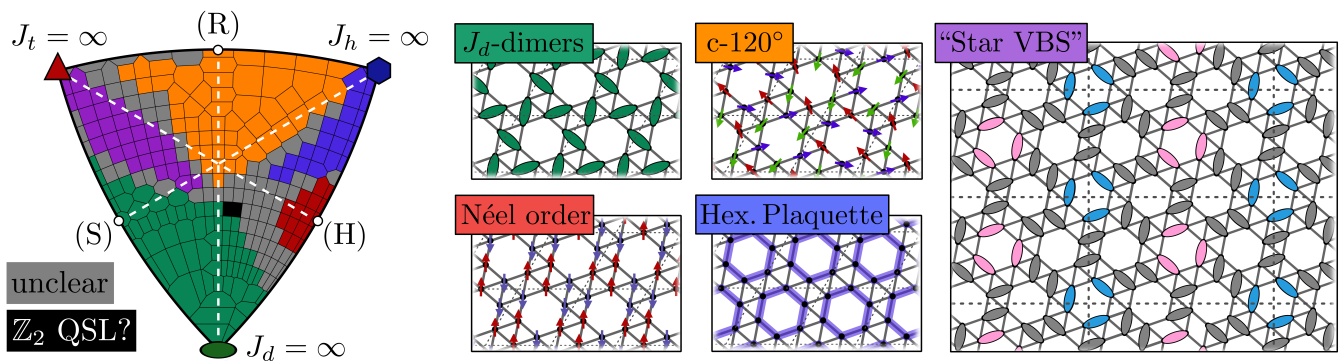}
        \caption{\label{fig:phaseDiagram}
            Phase diagram of the HAF~\eqref{eq:H} on the MLL based on $N=36$ ED data (Voronoi tesselation), where $0 \leq J_t, J_h, J_d \leq \infty$ (AFM octant on the $J_t^2 + J_h^2 + J_d^2=\mathrm{const.}$ sphere).
            The three external boundaries correspond to the Heisenberg models on the ruby (R), star (S) and honeycomb (H) lattices, respectively obtained by setting $J_d$, $J_h$ or $J_t$ to zero.
            Lines where two couplings coincide are drawn in white, marking the isotropic point $J_t=J_h=J_d$ at the center.
            Prototypical states for the color-coded phases are illustrated, including a novel ``star VBS'' state on $J_t, J_h$ bonds where the dimer motives on $J_h$-hexagons can resonate in two variations (light blue and pink, a 36-site pattern being shown here).
            Two proximate points, called P1 and P2 below, where Gutzwiller-projected gapped $\mathbb{Z}_2$ QSL ans\"atze are in strong agreement with the ED ground state, are drawn in black.
        }
        
    \end{figure*}

    In this work, we present the $J_t-J_h-J_d$ phase diagram of the HAF on the MLL, including a rather exotic VBS and a plaquette state as well as a candidate region where $\mathbb{Z}_2$ gapped spin-liquid ans\"atze are in astonishing agreement with the exact $N=36$ ground state.
    Furthermore, our MLL framework offers a unified perspective on the HAFs on the ruby, star and honeycomb lattices.

\paragraph*{Model.}

    We consider the spin-1/2 Hamiltonian
    \begin{equation}\label{eq:H}
        H = 
        J_t \sum_{\langle i, j\rangle \in t}   \bm{S}_i \cdot \bm{S}_j 
        + J_h \sum_{\langle i, j\rangle \in h}   \bm{S}_i \cdot \bm{S}_j
        + J_d \sum_{\langle i, j\rangle \in d}   \bm{S}_i \cdot \bm{S}_j,
    \end{equation}
    with the three NN couplings $J_t, J_h, J_d \geq 0$ as shown in Fig.~\ref{fig:lattice} and periodic boundary conditions on finite clusters.  
    Using the spherical parametrization
    \begin{align}\label{eq:sphericalParametrization}
        &J_t = \cos\phi\sin\theta,&
        &J_h = \sin\phi\sin\theta,&
        &J_d = \cos\theta,&
    \end{align}
    the purely antiferromagnetic (AFM) octant corresponds to $\phi, \theta \in [0, \pi/2]$.
    As on the Shastry-Sutherland lattice, the product state of $J_d$-dimer singlets is an exact eigenstate for any $J_t, J_h, J_d$, and is provably the ground state if $J_d/2 > J_t = J_h > 0$~\cite{ghosh:2022, ghosh:2025b}.
    If exactly one of the three bond types is neglected, the NN graph of
    the \emph{ruby lattice} for $J_d = 0$,
    the \emph{star lattice} for $J_h = 0$,
    or the \emph{honeycomb lattice} for $J_t = 0$ are obtained.
    These special cases constitute the three boundary lines of the AFM phase diagram, as the ratio of the two remaining bonds can be varied.
    The isotropic points on the ruby, star and honeycomb lattices are respectively marked by (R), (S), and (H) in Fig.~\ref{fig:phaseDiagram}.
    At the three corners of the phase diagram only one of the three bonds is active and the system forms isolated groups of three ($J_t$ triangles), six ($J_h$ hexagons), or two ($J_d$ ``dimers'') spins.
    
    For the isotropic HAF ($J_t = J_h$) on the ruby lattice, Refs.~\cite{farnell:2011, farnell:2014, richter:2004, nyckees:2025} see indications of a \canted{} order, while some studies report three-fold degenerate lattice-symmetry breaking VBS states~\cite{jahromi:2020} or paramagnetic ground states preserving the lattice symmetry~\cite{schmoll:2025}.
        
    The HAF on the star lattice is known to host the $J_d$-dimer ground state if $J_d \gg J_t$ while conflicting reports exist on its $J_d \ll J_t$ behavior~\cite{jahromi:2020, ghosh:2025a, yang:2010, misguich:2007, ishikawa:2024}.
    Recently Ref.~\cite{ghosh:2025a} argued that a ``$\sqrt{3}\times\sqrt{3}$ VBS'' is the likely ground state below $J_d/J_t\approx 0.18$.
        
    On the honeycomb lattice the isotropic ($J_h = J_d$) HAF is known to host a long-range-ordered (LRO) ground state with non-zero staggered magnetization, although significantly reduced by quantum corrections~\cite{reger:1989, krueger:2000, castro:2006}.

\paragraph*{Method.}

     \begin{figure*}
        \centering
        \includegraphics[width=0.9\textwidth]{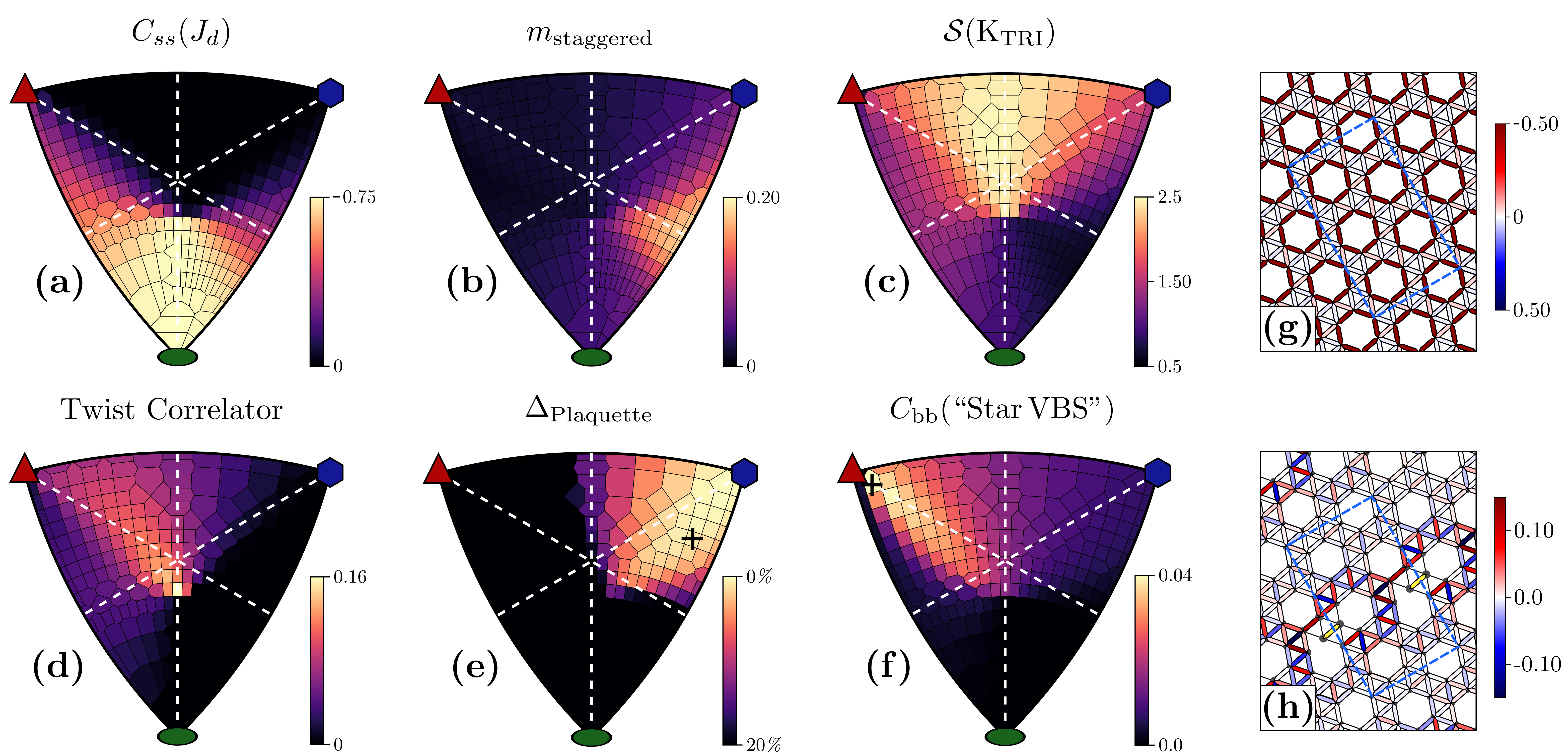}
        \caption{\label{fig:correlators-N36}
            ED ground state diagnostics on the $N=36$ cluster.
            (a): Average spin-spin correlator $C_{ss}(J_d)$ on $J_d$-bonds;
            (b): staggered magnetization per site;
            (c): (equal time) spin-spin structure factor $\mathcal{S}$ at the K point, $\mathrm{K}_\mathrm{TRI}$, of the underlying triangular lattice;
            (d): bond-averaged ``twist correlator'' $\langle D_{01} D_{ij} \rangle$ (see Eq.~\eqref{eq:Dij});
            (e): relative deviation of NN spin-spin correlator on $J_h$-bonds from their plaquette value (see Eq.~\eqref{eq:DeltaPlaquette});
            (f): connected bond-bond correlator $C_{bb}$ (see Eq.~\eqref{eq:Cbb}) for the ``star VBS'';
            (g): NN spin-spin correlator at the point marked in (e);
            (h): connected bond-bond correlator at the point marked in (f) with reference bond $\langle 0,1\rangle$ shown in yellow. 
        }
    \end{figure*}

\begin{figure}
        \centering
        \includegraphics[width=0.95\linewidth]{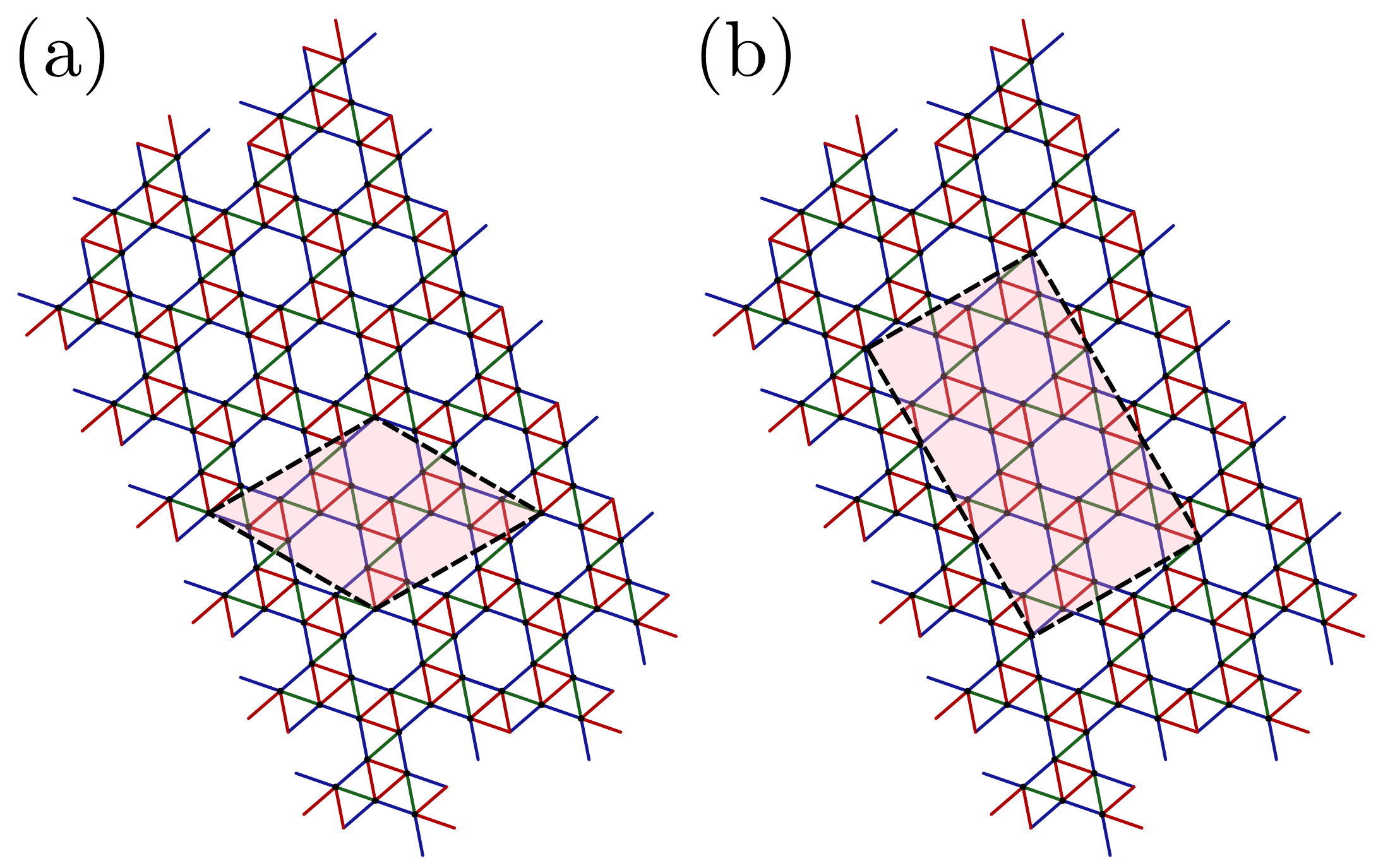}
        \caption{\label{fig:EndMatter:clusters}
        The two finite clusters used in all ED computations as rhombuses on the MLL; (a) for $N=18$ and (b) for $N=36$ spins.
        The same color-coding as in Fig.~\ref{fig:lattice} is assumed.
        }
    \end{figure}

    We performed extensive ED computations 
    using the XDiag library~\cite{wietek:2025:xdiag}, which fully resolves irreducible representations (irreps) of the space group as well as different $S^z_{\mathrm{tot}} = \sum_i S_i^z$ values.
    Out of all finite clusters viable for ED phase diagram scans only one $N=18$ and one $N=36$ cluster (see Fig.~\ref{fig:EndMatter:clusters} in End Matter) capture high-symmetry K points, crucial for ground state candidates such as the \canted{} order~\cite{farnell:2011, schmalfuss:2002, schulenburg:2000}.
    For a thorough explanation on how to obtain Gutzwiller-projected QSL ans\"atze we refer to the Supplemental Material (SM)~\cite{SupplemantalMaterial}.

\paragraph*{Phase Diagram.}

    Space group and $\mathrm{SU}(2)$ irreps of the lowest energy levels of Heisenberg models usually remain locally constant throughout the phase diagram, changing only at phase boundaries.
    We classified five contiguous parameter regions throughout which properties of ED spectra as well as (ground state) correlators and spin-spin structure factors remain uniform.
    Combining these diagnostics, we identify an extended $J_d$-dimer, a collinear N\'eel-order, the \canted{}, and a hexagonal plaquette phase as well as a novel VBS regime.
    The approximate extent of these phases on the AFM octant of the $J_t-J_h-J_d$ sphere is shown in Fig.~\ref{fig:phaseDiagram}.

\begin{figure}
        \centering
        \includegraphics[width=\linewidth]{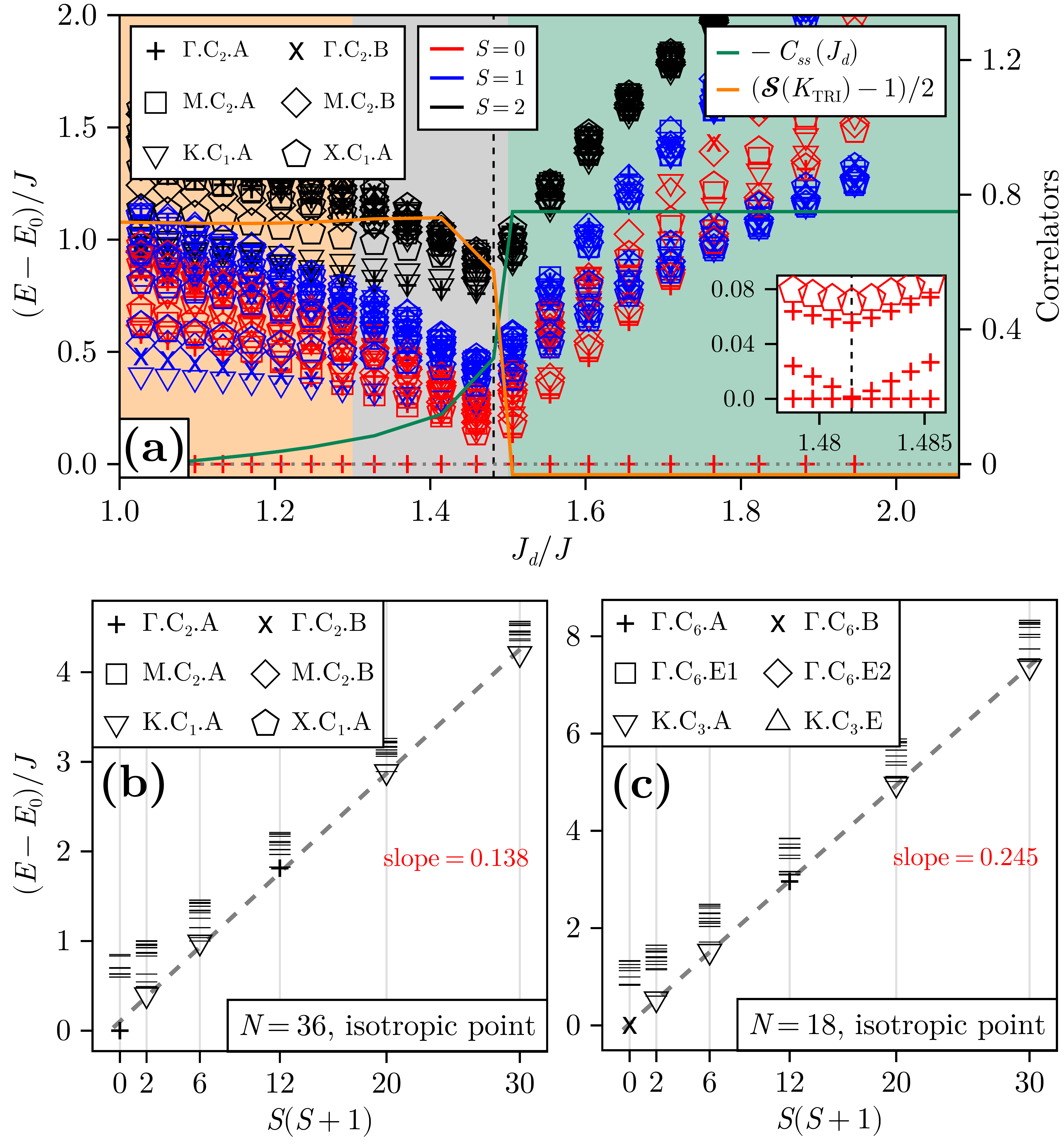}
        \caption{
        (a): Lowest eigenstates near the transition into the $J_d$-dimer phase along the $J=J_t=J_h$ line (vertical dashed line in the phase diagram from Fig.~\ref{fig:phaseDiagram}).
        The inset shows the first singlet excitations as the gap closes.
        (b) and (c): Tower of states of the $N=36$ and $N=18$ clusters at the isotropic point $J_t = J_h = J_d = J$.
        Dashed lines show the least-squares fit to the lowest eigenstate for each $S$.
        Other levels are merely drawn as lines for clarity.
        The energies of the lowest eigenstate for each $S$ are found to scale as $S(S+1)/N$ in agreement with a magnetically-ordered system.
        }
        \label{fig:TOS-isotropic}
    \end{figure}

    Starting on the vertical $J=J_t = J_h$ line connecting the isolated $J_d$-dimer point (bottom corner) to the ruby lattice point (R) on top, our results qualitatively agree with previous studies reporting LRO~\cite{farnell:2011, schmalfuss:2002, schulenburg:2000, gresista:2023, gembe:2024, beck:2024, nyckees:2025}, as we find structure factors compatible with the \canted{} order (\cantCol{} region) for $J_d/J < 1.48$ (also see Fig.~\ref{fig:TOS-isotropic}{a}) and strong $J_d$-dimerization (\JdCol{} region) above $1.5 \lesssim J_d/J$.
    Off the vertical $J_t = J_h$ line, the $J_d$-dimer phase is destabilized by $J_h$ faster than by $J_t$ interactions.
    This is consistent with Ref.~\cite{ghosh:2025a} locating the transition out of the $J_d$-dimer phase on the star lattice around $J_d/J_t\approx 0.18$.
    As expected, we find a collinear N\'eel phase at the isotropic honeycomb point (H)~\cite{reger:1989, krueger:2000, castro:2006}, which significantly extends into the MLL phase diagram (\neelCol{} region).
    The maple-leaf HAF thus hosts an exact dimer eigenstate next to a collinear N\'eel-ordered region, just as in the Shastry-Sutherland HAF~\cite{wietek:2019, yang:2022}.
    As $J_d/J_h$ is lowered further along the $J_t=0$ line, the model becomes a collection of $J_h$-hexagons that are weakly coupled by $J_d$-bonds.
    The overall ground state thus becomes a (dressed) product of ground states of the isolated 6-spin $J_h$-hexagons, which we call ``hexagonal plaquette'' state (also ``hexagonal singlet'' in Refs.~\cite{gresista:2025, ghosh:2024b, ghosh:2024a} where partially ferromagnetic couplings were considered).
    Its extent is shown by the \plaquetteCol{} region in Fig.~\ref{fig:phaseDiagram} which is consistent with previous triplon analysis and quantum Monte Carlo studies~\cite{adhikary:2021}.
    Lastly, we report a novel VBS state, in the \JtJhVBSCol{} region of Fig.~\ref{fig:phaseDiagram} in between the $J_d$-dimer and \canted{} regimes.
    As further discussed below, the ED spectra throughout this region suggest a ground state that significantly breaks the space group of both the $N=18$ and $N=36$ clusters while retaining spin-rotation symmetry.
    Indeed, bond-bond correlators in (h) of Fig.~\ref{fig:correlators-N36} hint at at a VBS pattern as the one illustrated to the right in Fig.~\ref{fig:phaseDiagram}, whose degree of freedom on hexagonal motives can explain the broken space-group symmetries.
    We refer to this state as the ``star VBS'' due to the dimers appearing to respectively form six rays around each hexagonal motive.
    The symmetry analysis of this ``star VBS'' performed below precisely predicts the irrep structure observed in our ED spectra on the $N=36$ cluster.

\paragraph*{Correlators and Structure Factors.}

    Spin-spin correlations show the extent of the $J_d$-dimer and collinear N\'eel regimes in (a) and (b) of Fig.~\ref{fig:correlators-N36} which coincide with the regions classified based on similarities in the eigenspectra.
    To estimate the extent of the \canted{} phase in (c), we use the fact that its spin-spin structure factor is concentrated at the high-symmetry K points, $K_{\mathrm{TRI}}$, of the underlying (undepleted) triangular lattice~\cite{gembe:2024, beck:2024, gresista:2023, schaefer:2025, gresista:2025}.
    Next we consider the following ``twist operator''
    \begin{equation}\label{eq:Dij}
        D_{ij} = (\hat{\bm{S}}_i \times \hat{\bm{S}}_j)^z = \hat{S}^x_i\hat{S}^y_j - \hat{S}^y_i\hat{S}^x_j.
    \end{equation}
    After assigning a suitable orientation to every bond $\langle i, j\rangle$, the ``twist correlator'' $\langle D_{01}D_{ij}\rangle$ relative to a reference bond $\langle 0, 1 \rangle$ can identify the uniform $120^{\circ}$(1) order~\footnote{
        There are two (typically degenerate) versions of the uniform (non-canted) $120^{\circ}$ order on a triangular lattice, sometimes called $120^{\circ}$(1) and $120^{\circ}$(2), which can be discriminated using the ``twist correlator''.
        For our choice of bond orientations $\langle i,j\rangle$, only the $120^{\circ}$(1) order yields a positive value for all $\langle D_{01}D_{ij}\rangle$, which is what we observe in subplots (d) of Fig.~\ref{fig:correlators-N36} and Fig.~\ref{fig:N-18-data}.
        The $120^{\circ}$(2) order would lead to a specific pattern of positive and negative values among all NN bonds, which is not observed.
    }.
    As signaled by the ``twist correlator'' around $J_d/J\approx 1.4$ in (d), our data reinforces the picture from Refs.~\cite{farnell:2011, nyckees:2025} that the canting angle in the \canted{} phase approaches the uniform $120^{\circ}$ order close to the $J_d$-dimer regime.
    To quantify plaquette-type spin-spin correlations (such as (g) in Fig.~\ref{fig:correlators-N36}) throughout the phase diagram, we compare NN spin-spin correlations on $J_h$-bonds to their plaquette value $e_{\text{Plaquette}}/J_h \approx -0.46713$ (the ground state energy per site of a six-spin NN Heisenberg AFM chain) via
    \begin{equation}\label{eq:DeltaPlaquette}
        \Delta_{\text{Plaquette}} = \frac{1}{N} \sum_{\langle i, j \rangle\in h} \frac{|\langle \hat{\bm{S}}_i \cdot \hat{\bm{S}}_j\rangle -e_{\text{Plaquette}}|}{|e_{\text{Plaquette}}|}.
    \end{equation}
    This quantity equals zero in the exact plaquette state and is shown in (e).
    Subfigure (f) displays the connected bond-bond correlator
    \begin{equation}\label{eq:Cbb}
        C_{\mathrm{bb}}(B) = \frac{1}{|B|} \sum_{\langle i, j \rangle \in B} \langle (\hat{\bm{S}}_0 \cdot \hat{\bm{S}}_1) (\hat{\bm{S}}_i \cdot \hat{\bm{S}}_j)  \rangle_{c}
    \end{equation}
    is shown, where $B$ is the set of NN bonds participating in the 36-site ``star VBS''~\footnote{
        Only NN bonds $\langle i, j \rangle$ that are not adjacent to the reference bond $\langle 0,1\rangle$ enter the set $B$ in $C_{\mathrm{bb}}(B)$.
    } shown Fig.~\ref{fig:phaseDiagram}.
    A real-space depiction of the underlying bond-bond correlator is shown in (h).
    Remarkably, it displays two clearly positively (red) and four rather negatively (white to blue) correlated hexagons, the latter being connected by red parallel bonds mirroring the grey bonds in Fig.~\ref{fig:phaseDiagram}.
    Correlators and spectra on the $N=18$ cluster as well as comments on the ``$\sqrt{3}\times\sqrt{3}$ VBS'' that was reported to be the ground state on the star lattice for $J_d \ll J_t$ ~\cite{ghosh:2025a} can be found in SM~\cite{SupplemantalMaterial}.

\paragraph*{Energy Spectroscopy.}

\begin{figure*}[!t]
        \centering
        \includegraphics[width=\textwidth]{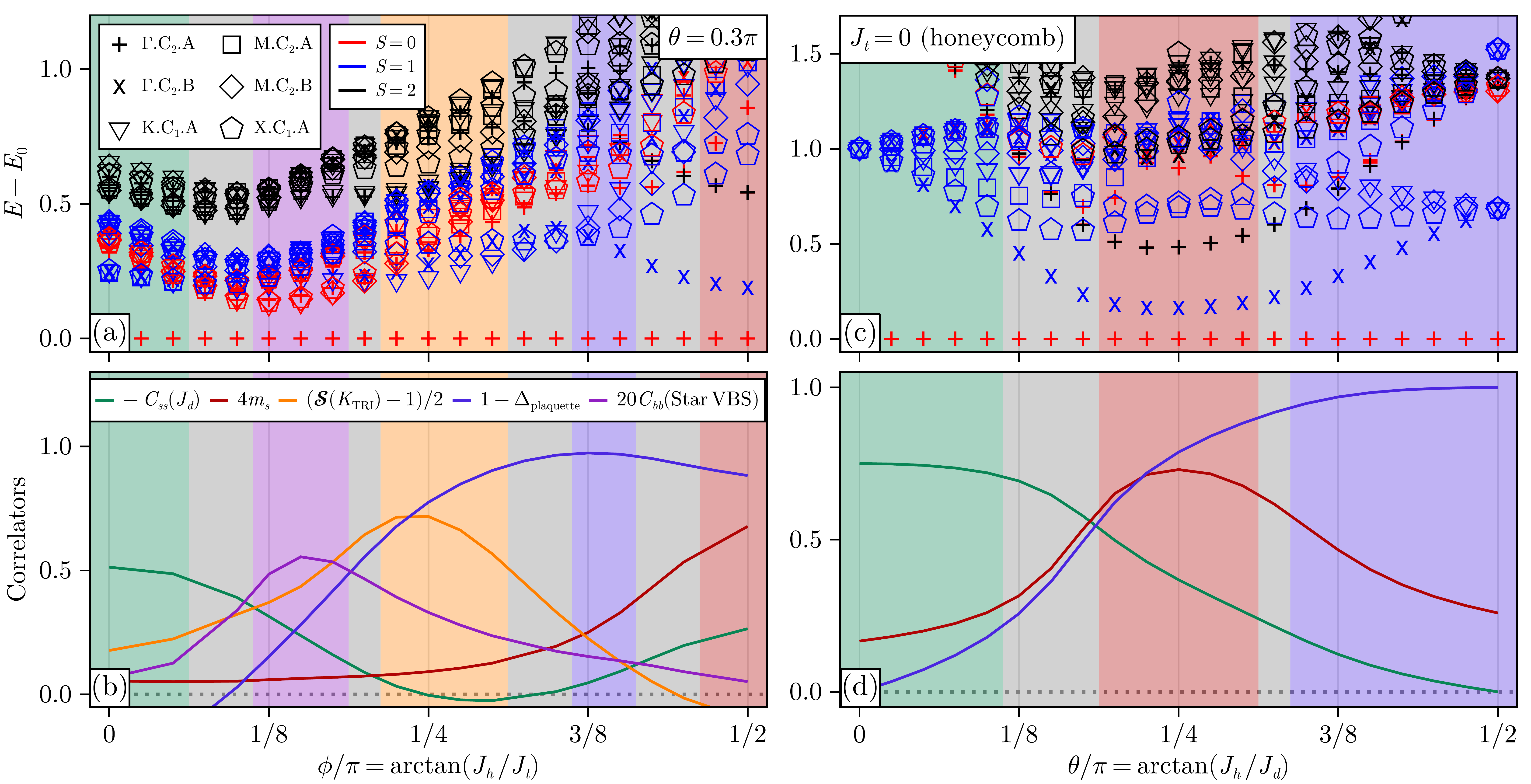}
        \caption{\label{fig:app:N36-cuts}
            (a) and (b): Lowest eigenstates and correlators along the (horizontal) $\theta=0.3\pi$ cut through the phase diagram in Fig.~\ref{fig:phaseDiagram}.
            (c) and (d): Lowest eigenstates and correlators along the $\phi=\pi/2$ $(J_t = 0)$ cut corresponding to the honeycomb HAF where every second bond in every second hexagon is varied.
            Background colors show the phases associated with different parameter regions on the $N=36$ cluster and match Fig.~\ref{fig:phaseDiagram}.
            In contrast to (a) in Fig.~\ref{fig:TOS-isotropic}, the overall bond strength is normalized such that $J_t^2 +J_h^2 + J_d^2 = 1$ throughout the phase diagram.
        }
    \end{figure*}

    $\mathrm{SU}(2)$ and space-group irreps of the lowest-lying states are a central tool through which we classify the extent of different phases in Fig.~\ref{fig:phaseDiagram}.
    As a starting point, we show part of the $J=J_t=J_h$ line near the transition into the $J_d$-dimer phase in (a) of Fig.~\ref{fig:TOS-isotropic}, featuring singlet ($S=0$), triplet ($S=1$), and quintuplet ($S=2$) levels as well as all space-group irreps (see \footnote{
        Our short-hand notation of little-group irreps is defined as follows: For a given momentum $\bm{k}$, the little group $\mathcal{G}(\bm{k})$ contains all space-group operations that preserve $\bm{k}$ and is a space group itself. The little co-group $\hat{\mathcal{G}}(\bm{k})$ is the point group of said little group and (at least for symmorphic lattices such as the MLL) an irrep $\rho$ of $\hat{\mathcal{G}}(\bm{k})$ uniquely determines an irrep of $\mathcal{G}(\bm{k})$ which (together with the star of $\bm{k}$) uniquely determines an irrep of the original space group, which we refer to as $\bm{k}.\hat{\mathcal{G}}(\bm{k}).\rho$.
        For instance, $\Gamma$.C$_6$.B refers to the space-group irrep induced by the irrep of $\mathcal{G}(\Gamma)$ which was induced from the B irrep of its point group $\hat{\mathcal{G}}(\Gamma)=\text{C}_6$, where C$_n$ denotes the cyclic group of order $n$.
    } for our notation) of the $N=36$ cluster.
    The average spin-spin correlator on $J_d$-bonds, $C_{ss}(J_d)$, and the aforementioned structure factor, $\mathcal{S}(\mathrm{K}_\mathrm{TRI})$, are drawn as solid lines.
    The $J_d$-dimer phase shows a strong separation of different $S$-levels, while the \canted{} region displays a characteristic K.C$_1$.A first excited $S=1$ triplet followed by a  K.C$_1$.A first $S=2$ quintuplet.
    A gap-closing is observed at $J_d/J \approx 1.48$, while the first excited levels become pure singlets in the range $1.3 \lesssim J_d/J \lesssim 1.7$.
    This could be due to finite-size cross-overs or be related to previously hypothesized intermediate spin-liquid phases~\cite{gresista:2023}.
    As can be seen based on the spectra in Fig.~\ref{fig:app:N36-cuts} in the End Matter, the N\'eel phase is cleanly characterized by the first and second excited states respectively being a $\Gamma$.C$_2$.B triplet and $\Gamma$.C$_2$.A quintuplet.
    Most strikingly, the region in which the correlator (f) from Fig.~\ref{fig:correlators-N36} lights up is accompanied by singlet first-excited levels on both system sizes, the significance of which is discussed below.

\paragraph*{Tower of States Analysis.}

    For magnetically-ordered phases of $\mathrm{SU}(2)$-symmetric Hamiltonians it is known that the energy of the lowest spin-$S$ state scales as $S(S+1)/N$~\cite{wietek:2016}.
    Precisely this behavior is observed at the isotropic point $J_t=J_h=J_d$ in (b) and (c) of Fig.~\ref{fig:TOS-isotropic}, strongly suggesting that the \canted{} region does indeed magnetically order on the $N=18,36$ clusters.
    These tower of states (TOS) plots are consistent with previous studies~\cite{schulenburg:2000, richter:2004} that were not fully resolving all space-group irreps.
    Likewise, the TOS in the N\'eel-ordered region near the honeycomb limit collapses into a $\Gamma$.C$_6$.A, $\Gamma$.C$_6$.B-degenerate ground state as $N\to\infty$.
    
    Importantly, if the system realizes a lattice-symmetry breaking state in the thermodynamic limit, one expects an $S=0$ ground state with space-group irrep degeneracy~\footnote{
        A lattice-symmetry breaking state can be fit onto the lattice in multiple ways, all of which being related by a space group and a trivial/identity $(S=0)$ spin operation.
    }.
    On finite clusters such degeneracies are usually lifted, leading to groups of $S=0$ first-excited states, just as observed within the \JtJhVBSCol{} region in Fig.~\ref{fig:phaseDiagram}.
    With methods described in the SM~\cite{SupplemantalMaterial}, it is possible to compute the (super) set of space-group irreps that can appear as degenerate ground states of a prototypical state in the thermodynamic limit.
    A list of these irreps for relevant states in the HAF on the MLL can be found in Tab.~\ref{tab:multiplicities} in the SM~\cite{SupplemantalMaterial}, where we e.g. obtain a $\Gamma$.C$_6$.A$,\Gamma$.C$_6$.B-degenerate ground state for the collinear N\'eel phase or a $\Gamma$.C$_6$.A, $\Gamma$.C$_6$.B, K.C$_3$.A degeneracy for the \canted{} phase in the thermodynamic limit.
    Both of these are consistent with our data and in particular the TOS shown in Fig.~\ref{fig:TOS-isotropic}, taking into account that the $N=36$ cluster does only resolve the little co-groups $\Gamma$.C$_2$ and K.C$_1$ instead of $\Gamma$.C$_6$ and K.C$_3$.
    Most astonishingly, if the ground state irreps predicted for the 36-site ``star VBS'' in the thermodynamic limit are translated to the lower-symmetry $N=36$ cluster, we precisely obtain $\Gamma$.C$_2$.A, K.C$_1$.A, M.C$_2$.B, and X.C$_1$.A, i.e., the singlet excitations observed throughout the \JtJhVBSCol{} region of the phase diagram in Fig.~\ref{fig:phaseDiagram}.

    \paragraph*{Variational wave-functions for putative QSL region.} 
    A complete classification of mean-field Ans\"atze for fully symmetric QSLs has been accomplished in Ref.~\cite{sonnenschein:2024}.
    In Tab.~\ref{tab:QSL-Data} we report variational energies of corresponding Gutzwiller-projected wave-functions obtained from variational Monte Carlo at the two points drawn in black in the phase diagram of Fig.~\ref{fig:phaseDiagram}.
    Among the twelve $\mathrm{U}(1)$ and eight $\mathbb{Z}_{2}$ states realized with NN mean-field amplitudes, we find two specific {\it gapped} $\mathbb{Z}_{2}$ QSLs (Z0002 and Z1102 in the nomenclature of Ref.~\cite{sonnenschein:2024}) yielding competitive energies to the ED ground state on the same 36-spin cluster.
    The accuracy of these variational energies is at the order of $|(E_{\rm VMC}-E_{\rm ED})/E_{\rm ED}|\sim10^{-3}$, i.e., an order of magnitude smaller compared to the U(1) Dirac spin-liquids on the kagome and triangular lattices~\cite{Iqbal-2013,Iqbal-2016}, both known to provide an excellent variational description of the ground state of the respective spin-1/2 HAF~\cite{Wietek-2024}.
    We find that this QSL is stable only when $J_{t}\neq J_{h}$ while for $J_{t}=J_{h}$ they destabilize to the $J_d$-dimer product state.
    The impressive agreement of spin-spin correlations between ED and the two $\mathbb{Z}_2$ QSL Ans\"atze is shown in the End Matter.
    \begin{table}
        \centering
        \caption{\label{tab:QSL-Data}
                ED ground state and variational energies per site of corresponding Gutzwiller-projected wave-functions of the Z0001 and Z1102 $\mathbb{Z}_2$ QSL Ans\"atze~\cite{sonnenschein:2024} on different clusters of $N$ sites (the $N=216,384$ clusters respect all symmetries of MLL).
                The two points P1 and P2 are marked in black in the phase diagram of Fig.~\ref{fig:phaseDiagram} at the exact coordinates $\theta/\pi = 0.225$, $\phi/\pi = 0.275$ and $\theta/\pi = 0.225$, $\phi/\pi = 0.3$.
            }
        \begin{tblr}{width=0.7\linewidth,
                    hlines=0.4pt,
                    vlines =0.4pt,
                    colspec = {cllc},
                    colsep = 2pt,
                    rowsep = 1pt,
                    hline{1} = {0pt},
                    hline{11} = {0pt},
                    vline{1} = {0pt},
                    vline{5} = {0pt},
                    }
                           &\SetCell[r=1]{c}P1           &\SetCell[r=1]{c}P2&$N$\\
                $J_h/J_t$  &   \SetCell[r=1]{c}$1.171$   & \SetCell[r=1]{c}$1.376$ &  \\  
                $J_d/J_t$  &   \SetCell[r=1]{c}$1.803$   & \SetCell[r=1]{c}$1.992$ & \\  
                $E_\mathrm{ED}/J_t$  &   $-0.6802702$    & $-0.7652702$ & 36\\           
                $E^{Z0002}_\mathrm{VMC}/J_t$ & $-0.679082(1)$ & $-0.762053(2)$ & 36\\
                $E^{Z1102}_\mathrm{VMC}/J_t$ & $-0.679051(1)$ & $-0.761414(2)$ & 36 \\
                $E^{Z0002}_\mathrm{VMC}/J_t$ & $-0.679242(2)$ & $-0.762807(3)$ & 216 \\
                $E^{Z1102}_\mathrm{VMC}/J_t$ & $-0.679089(2)$  & $-0.762206(3)$  & 216 \\
                $E^{Z0002}_\mathrm{VMC}/J_t$ & $-0.67924(1)$ & $-0.76280(1)$ & 384 \\
                $E^{Z1102}_\mathrm{VMC}/J_t$ & $-0.67911(1)$ & $-0.76223(2)$  & 384 
            \end{tblr}
        \end{table}

\paragraph*{Discussion.}
Studying the phase diagram of the maple-leaf antiferromagnet using exact diagonalization, we discovered a valence bond solid state with hexagonal motifs in the vicinity of an extended (canted) $120^\circ$ N\'eel state.
The transition between the $120^\circ$ N\'eel state and the VBS is strongly reminiscent of the instabilities of the Dirac spin-liquid on the triangular lattice, as relevant for the $J_1$-$J_2$ Heisenberg model.
There, signatures of a VBS with a twelve-site unit cell have also been found next to a $120^\circ$ N\'eel state~\cite{Wietek-2024}, where both of these states can be considered instabilities of the $\pi$-flux Dirac spin-liquid~\cite{Hermele2004,Hermele2005}. Remarkably, the VBS state extends over a large region of the phase diagram, almost ranging from the isolated-triangle limit to the isotropic point.
As such, we find that the state of isolated triangles is easily destabilized into the ``star VBS'' state.
In contrast, the state of isolated hexagons shows great stability, extending in the direction of the isotropic point and the exact $J_d$-dimer eigenstate.
The interplay between a collinear antiferromagnet, a plaquette phase, and an exact dimer product state is strongly reminiscent of the Shastry-Sutherland model, where a plaquette state was found to be located between a dimer product state and a collinear antiferromagnet~\cite{Corboz2013}. 
Furthermore, the presented phase diagram is consistent with recent experiments where a strong tendency for $J_d$-dimerization was found in Ho$_3$ScO$_6$, likely realizing a distorted maple-leaf lattice spin-2 model with $J_d \gg J_t > J_h > 0$~\cite{Aguilar-Maldonado:2025}.

We identify a region of the phase diagram in which the exact $J_d$-dimer eigenstate, the collinear N\'eel state, a (canted) $120^\circ$ N\'eel state and the hexagonal plaquette state all compete, presenting an ideal regime for the emergence of a putative quantum spin-liquid.
Indeed, by comparing variational energies of a broad class of gapless $\mathrm{U}(1)$ and gapped $\mathbb{Z}_2$ spin-liquids, we discover that two specific $\mathbb{Z}_2$ {\it Ansätze} exhibit a remarkably low variational energy compared to the exact $N=36$ ground state and qualitatively reproduce the exact spin-spin correlations.
As such, we identify a highly promising region for the emergence of the long sought-after $\mathbb{Z}_2$ quantum spin liquid with potentially exotic forms of quantum criticality governing the observed quantum phase transitions. 

\paragraph*{Acknowledgments.}
We thank Karlo Penc, Robin Sch\"afer, Rafael D. Soares, Philipp Schmoll, and Jan Naumann for very helpful discussions. A.W. acknowledges support by the German Research Foundation (DFG) through the Emmy Noether program (Grant No. 509755282), IIT Madras for a Visiting Faculty Fellow position under the IoE program, and the European
Research Council (ERC) under the European Union’s Horizon
Europe research and innovation program (Project ID 101220368)—ERC Starting Grant MoNiKa. Y.I. acknowledges support from the ICTP through the Associates Programme and from the Simons Foundation through Grant No. 284558FY19. Y.I. also acknowledges the use of the computing resources at
HPCE, IIT Madras.

\bibliography{main}
\clearpage

\onecolumngrid
\begin{center}
\textbf{\large End Matter}
\end{center}
\twocolumngrid

\paragraph*{Appendix A: Finite Clusters.}

    Due to the computational complexity of ED, our phase diagram scans are restricted to rather small system sizes.
    The only two clusters of manageable size that resolve high-symmetry K points~\cite{farnell:2011, schmalfuss:2002, schulenburg:2000} consist of $N=18$ and $N=36$ spins and are shown in Fig.~\ref{fig:EndMatter:clusters}.
    The next K-resolving cluster consists of $N=54$ spins.
    While the $N=18$ cluster is highly symmetric, resolving $\Gamma$ and K with their little co-groups C$_6$ and C$_3$, the $N=36$ system only has an inherent C$_2$ symmetry but captures $\Gamma$, K, X, and M points, albeit not their full little co-groups.
    Both clusters have been used in previous studies~\cite{farnell:2011, schmalfuss:2002, richter:2004}.

\paragraph*{Appendix B: $N=36$ Phase Diagram Cuts.}

    In Fig.~\ref{fig:app:N36-cuts} we present two more energy spectra for cuts through the phase diagram shown in Fig.~\ref{fig:phaseDiagram} in the main text:
    
    Subplots (a) and (b) present a ``horizontal'' cut at a moderate $J_d\approx 0.59$ ($\theta=0.3\pi$ in Eq.~\eqref{eq:sphericalParametrization}) from a star lattice limit with $J_d/J_t = \cot \theta \approx 0.73$ on the left to a honeycomb limit on the right with $J_d/J_h = \cot \theta \approx 0.73$.
    For small $\phi$ values, we still find relatively large spin-spin correlations on $J_d$-bonds (\JdCol{} solid line) combined with a $J_d$-dimer-like excitation spectrum.
    This is consistent with Ref.~\cite{ghosh:2025a} identifying the transition out of the $J_d$-dimer phase on the star lattice only around $J_d/J_t\approx 0.18$.
    As $\phi$ is further increased, we cut through the regime where the first four excitations are $S=0$ singlets, namely $\Gamma$.C$_2$.A, K.C$_1$.A, M.C$_2$.B, and X.C$_1$.A.
    The onset of singlet excitations nicely fits the peak in the connected bond-bond correlator of the 36-site ``star VBS'' (\JtJhVBSCol{} solid line) which does significantly break lattice symmetries and is expected to have a ground state of degenerate space-group irreps in the thermodynamic limit.
    Around the $\phi/\pi=1/4$ mark (note the proximity to the isotropic point at $\phi/\pi=1/4$ and $\theta/\pi = \arctan(\sqrt{2})/\pi \approx 0.30409$) we start to see the characteristic excitation spectrum of the \canted{} region, where the K.C$_1$.A irrep constitutes both the first triplet and quintuplet excitation, the former being the first excited state.
    This is accompanied by a strong signal of the spin-spin structure factor $\mathcal{S}(\mathrm{K}_\mathrm{TRI})$ (\cantCol{} solid line), which was observed as an important feature of this phase in Refs.~\cite{beck:2024, gresista:2023, gembe:2024, schaefer:2025, gresista:2025}.
    Hereafter, we encounter a small ``hexagonal plaquette'' region with $J_h>J_d>J_t$ before the N\'eel phase can be stabilized as $J_t \to 0$ and $J_d/J_h \to 0.73$, the latter being a notable deviation from the isotropic honeycomb HAF for which the collinear N\'eel phase is known~\cite{reger:1989, krueger:2000, castro:2006}.

    Subplots (c) and (d) in Fig.~\ref{fig:app:N36-cuts} show the honeycomb $(J_t = 0)$ limit from the isolated $J_d$-dimer (left) to the isolated $J_h$-hexagon limit (right).
    As hinted at by the previous cut, we see that the (red) collinear N\'eel phase is remarkably stable on the $N=36$ cluster and has a characteristic $\Gamma$.C$_2$.B triplet and a $\Gamma$.C$_2$.A quintuplet as first and second excited states.
    Along this cut our phase diagram is consistent with Ref.~\cite{adhikary:2021} which employs triplon analysis and quantum Monte Carlo techniques.
    For instance, their data implies the collinear N\'eel phase to be present between $0.2 \lesssim \theta/\pi \lesssim 0.33$, which is in agreement with the region between $0.2 \lesssim \theta/\pi \lesssim 0.3$ that we are able to unambiguously characterize based on the spectra shown in (c).

\paragraph*{Appendix C: QSL Candidate Spin-Spin Correlations.}

    \begin{figure*}[!b]
        \centering
        \includegraphics[width=\linewidth]{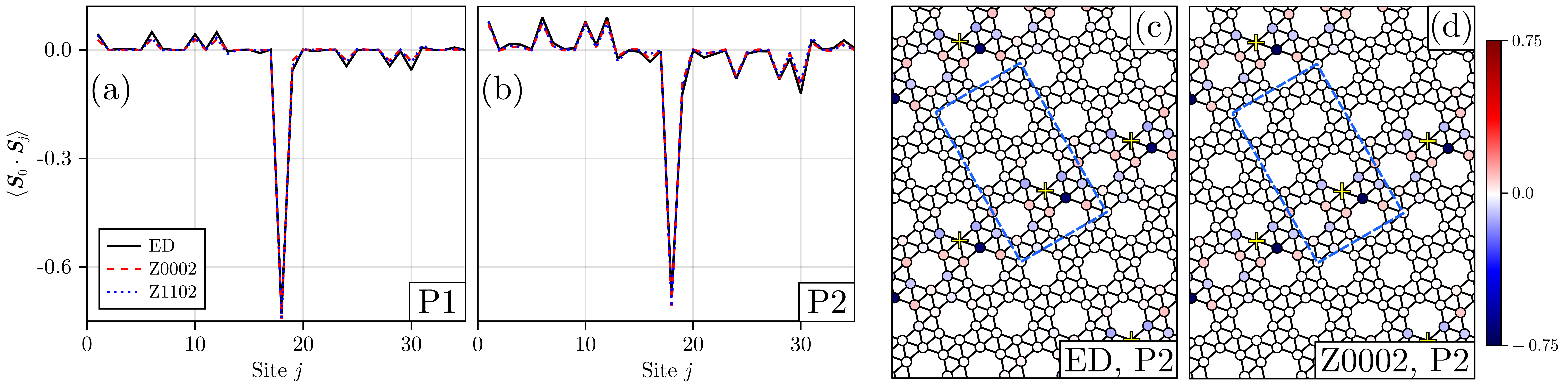}
        \caption{\label{fig:End:QSL-spin-spin}
            (a) and (b): Equal-time spin-spin correlations of the ED ground state and the two $\mathbb{Z}_2$ QSL Gutzwiller-projected wave-functions Z0002 and Z1102 on the $N=36$ cluster, respectively at the points P1 and P2 in the phase diagram.
            (c) and (d): Real-space depiction of the ED and Z0002 spin-spin correlation data from (b), both sharing the same color bar.
        }
    \end{figure*}

    As discussed in the main text, the energies of the Gutzwiller-projected wave-functions of the two gapped $\mathbb{Z}_2$ QSL Ans\"atze Z0002 and Z1102~\cite{sonnenschein:2024} on the $N=36$ cluster are astonishingly close to exact ground state at the two points P1 ($\theta/\pi = 0.225$, $\phi/\pi = 0.275$) and P2 ($\theta/\pi = 0.225$, $\phi/\pi = 0.3$).
    The Z0002 state, characterized by uniform hopping and $s$-wave pairing,
    has a smaller variance of the energy compared to the Z1102 state, and thus likely provides an excellent variational description of the true ground state~\cite{Gros-1990}.
    As can be seen in Fig.~\ref{fig:End:QSL-spin-spin}, both Ans\"atze reproduce the spin-spin correlations of the true ground state on the $N=36$ cluster at the points P1 and P2 with stunning accuracy.
    Furthermore, the exact as well as the Z0002 and Z1102 spin-spin correlations are short-ranged and focus mainly on satisfying the $J_d$-bond at P1 and P2, the latter being shown in (c) and (d) of Fig.~\ref{fig:End:QSL-spin-spin}.
    Although a certain degree of correlation is maintained on the $J_h$-hexagons, correlations generally appear to be unable to reach further than three neighboring sites.

    \onecolumngrid

    \clearpage
    \onecolumngrid          
    \appendix   
    \section*{Supplemental Material}
    \setcounter{page}{1}
\setcounter{section}{0}
\setcounter{figure}{0}
\setcounter{table}{0}
\setcounter{equation}{0}
\renewcommand{\thefigure}{S\arabic{figure}}
\renewcommand{\thetable}{S\arabic{table}}
\renewcommand{\theequation}{S\arabic{equation}} 
\renewcommand{\thepage}{S-\arabic{page}} 

    \section{Correlators and Spectra of the 18-Spin System}

        \begin{figure*}[!h]
            \centering
            \includegraphics[width=\textwidth]{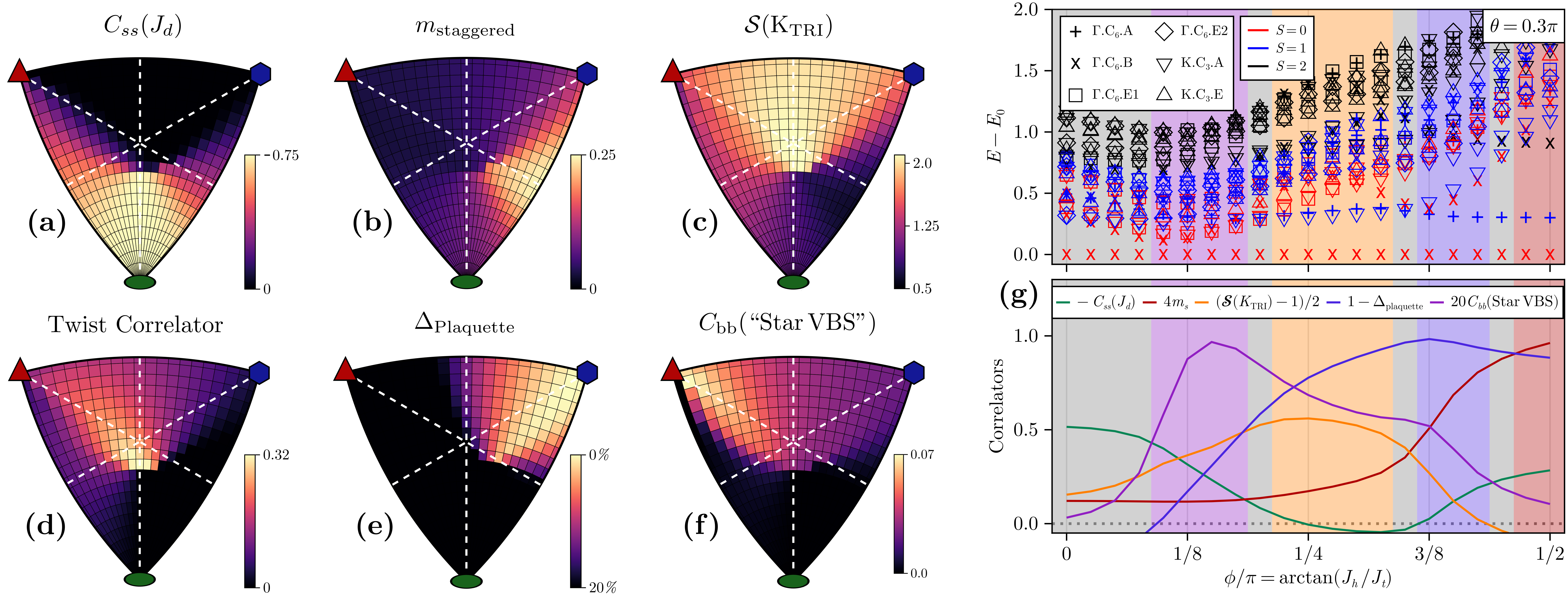}
            \caption{\label{fig:N-18-data}
                (a)-(f): $N=18$ ground state correlators and structure factors, similar to Fig.~\ref{fig:correlators-N36} in the main text.
                (g): ED spectra and correlator data on the $N=18$ system along the $\theta=0.3\pi$ cut in terms of the parametrization from Eq.~\eqref{eq:sphericalParametrization}.
                $\phi=0$ and $\phi=\pi/2$ correspond to the star lattice and honeycomb limits respectively, passing near the isotropic point at $\phi/\pi=1/4$.
            }
    \end{figure*}

        The phase diagram of the $N=18$ system looks similar to Fig.~\ref{fig:phaseDiagram}, although transitions are generally more ``washed-out'' due to the smaller system size.
        To demonstrate this, we show $N=18$ correlator data analogous to Fig.~\ref{fig:correlators-N36} from the main text in Fig.~\ref{fig:N-18-data} as well as the spectrum and correlator data along the $\theta = 0.3\pi$ cut already discussed in panel (a) of Fig.~\ref{fig:app:N36-cuts} for the $N=36$ system.
        
        In general, the behavior of every $N=36$ correlator is reproduced by the $N=18$ system, with the caveat that {$C_{bb}(\text{``Star\,VBS''})$} correlator in (f) of Fig.~\ref{fig:N-18-data} now corresponds to the 18-site version of the ``star VBS'' where one swaps all pink hexagonal motives in Fig.~\ref{fig:phaseDiagram} to blue (or vice versa; both correlators are essentially equal in the ground state).
        The $\theta = 0.3\pi$ cut in (g) is also very similar to the (a) and (b) cut in Fig.~\ref{fig:app:N36-cuts} for $N=36$, one difference being that the spectra close to the star lattice limit (left side in (g)) more clearly violate the degeneracy of space-group irreps at the same $S$-level.
        Due to this, we do not highlight the $\phi/\pi < 1/16$ region as $J_d$-dimerized compared to the $N=36$ cut (a) in Fig.~\ref{fig:app:N36-cuts}.
        Another difference worth mentioning is that the observed lowest $S=0$ levels in the \JtJhVBSCol{} region on the $N=18$ cluster are $\Gamma$.C$_6$.B, $\Gamma$.C$_6$.E1, and K.C$_3$.A, while the (super) set of irreps available to the 18-site ``star VBS'' state are $\Gamma$.C$_6$.A, $\Gamma$.C$_6$.B, and K.C$_3$.A according to Tab.~\ref{tab:multiplicities}.
        We thus summarize that the bond-bond correlator data is in good agreement with the 18-site ``star VBS'' being realized in the \JtJhVBSCol{} regime of the phase diagram of the $N=18$ cluster, while there remains an inconsistency in the $\Gamma$.C$_6$.E1 quantum number.

    \section{Comments on the Star Lattice VBS Ground State}

         Ref.~\cite{ghosh:2025a} analyzed multiple VBS states on the star lattice in the $J_d \ll J_t$ regime, finding a ``$\sqrt{3}\times\sqrt{3}$ (18-site) VBS'' state to be the likely ground state.
         We also computed connected bond-bond correlators with respect to this ``$\sqrt{3}\times\sqrt{3}$ VBS'', finding it to light up close to the $J_t=\infty$ point in the star lattice limit just below the dashed white line in Fig.~\ref{fig:phaseDiagram}.
         However, the signal is weaker than for the ``star VBS'' for both $N=18,36$ and rapidly decays away from the star lattice boundary of the phase diagram.
         It is thus plausible that the ``$\sqrt{3}\times\sqrt{3}$ VBS'' from Ref.~\cite{ghosh:2025a} is the ground state in the $J_d \ll J_t$ regime of the star lattice (the $C_{bb}(\text{``Star\,VBS''})$ correlator is zero along the whole star lattice line for $N=18,36$), but it does not appear to be competitive in the presence of $J_h$ interactions.
         Furthermore, a state with an 18-site magnetic unit cell does not resolve M or X points and thus cannot explain the M.C$_2$.B or X.C$_1$.A singlet excitations we see throughout the \JtJhVBSCol{} region in Fig.~\ref{fig:phaseDiagram} on the $N=36$ cluster.
         For the $N=18$ system, the observed singlet excitations in the \JtJhVBSCol{} region are $\Gamma$.C$_6$.B, $\Gamma$.C$_6$.E1, and K.C$_3$.A while the ``$\sqrt{3}\times\sqrt{3}$ VBS'' can only cause $\Gamma$.C$_6$.A or K.C$_3$.A to appear.
         The picture that the ``$\sqrt{3}\times\sqrt{3}$ VBS'' is realized in the \JtJhVBSCol{} region of the phase diagram on the $N=18$ cluster thus leads to one more inconsistent quantum number as compared to the 18-site ``star VBS''.

    \section{Space-Group Irreps of Selected States}
    
        To derive the space-group irreps associated with a certain set of periodic states in the thermodynamic limit, we start from a prototypical (product) state $|\psi\rangle$ on the infinite lattice, e.g., a specific up-down pattern if one is interested in collinear N\'eel states.
        In the case of VBS states, one might pick a classical spin arrangement with precisely the same symmetries.
        Then the space-group orbit of $|\psi\rangle$ is a representation of the space group by construction and can be decomposed into space-group irreps.
        Since for periodic states $|\psi\rangle$ the orbit must be finite, the representation matrices and their characters can be computed explicitly and the multiplicities appearing in the irrep decomposition be evaluated through ``scalar product of characters''-type formulae~\cite{wietek:2016}.
        There is a caveat, however, namely the states in the orbit will typically not be eigenstates and thus hybridize such that the lowest-energy states may only transform under a subset of the previously computed irreps.
    
        Tab.~\ref{tab:multiplicities} contains the potential space-group irreps of various states on the MLL.
        As discussed in the main text, these are consistent with the ED spectra observed in established phases such as the N\'eel~\cite{reger:1989, krueger:2000, castro:2006} or regime.
        Furthermore, the TOS plots at the isotropic point shown in panels (b) and (c) of Fig.~\ref{fig:TOS-isotropic} are consistent with the picture of a magnetically ordered \canted{} phase according to Tab.~\ref{tab:multiplicities}.
    
        \begin{table*}
        \centering
        \caption{\label{tab:multiplicities}Multiplicities of little group $\mathcal{G}(\bm{k})$ irreps contributing to the space-group orbit of certain states on the (infinite) MLL. 
            Recalling that C$_6$.E1, C$_6$.E2 and C$_3$.E are two-dimensional, the adjusted row sum yields the length of the semi-classical space-group orbit, i.e., the multiplicities already include the factor coming from the respective $\bm{k}$ star.
            Dashes are inserted whenever the magnetic unit-cell of a state is too small to resolve the respective $\bm{k}$-vector in which case there will be no Bloch-functions for $\bm{k}$ in the orbit, i.e., they can be thought of as zeros.}
        \begin{tblr}{width=\linewidth,
                    hlines=0.4pt, vlines=0.4pt,
                    colspec = {c *{9}{X[c,m]}},
                    colsep = 4pt,
                    rowsep = 2pt,
                    hline{1} = {1pt},
                    hline{2} = {1pt},
                    hline{3} = {1.5pt},
                    vline{1} = {1pt},
                    vline{2} = {1.5pt},
                    vline{6} = {1pt},
                    vline{8} = {1pt},
                    vline{10} = {1pt},
                    vline{11} = {1pt},
                    }
                \SetCell[r=2]{c}\textbf{Spin order}  &\SetCell[c=4]{c} $\Gamma$.C$_6$ &&&& \SetCell[c=2]{c} K.C$_3$  &&  \SetCell[c=2]{c} M.C$_2$ && X.C$_1$ \\
                                    &   A   &  B  &  E1  &   E2      &      A      &        E  &     A      &        B   &   A  \\
                $J_d$-dimers         &   1   &  0  &   0  &    0      &      -      &     -     &      -     &      -     &   -   \\                   
                N\'eel AFM          &   1   &  1  &   0  &    0      &      -      &     -     &      -     &      -     &   -  \\
                Hex. Plaquette      &   1   &  0  &   0  &    0      &      -      &     -     &      -     &      -     &   -  \\
                uniform $120^{\circ}$    &   1   &  1  &   0  &    0      &      4      &     0     &      -     &      -     &   -  \\
                canted $120^{\circ}$      &   1   &  1  &   0  &    0      &      4      &     0     &      -     &      -     &   -  \\
                $\sqrt{3}\times\sqrt{3}$ VBS &   1   &  0  &   0  &    0      &      2      &     0     &      -     &      -     &   -  \\
                18-site ``star VBS''      &   1   &  1  &   0  &    0      &      4      &     0     &      -     &      -     &   -  \\
                36-site ``star VBS''      &   1   &  0  &   0  &    1      &      2      &     2     &      0     &      3     &   6  \\
                
            \end{tblr}
        \end{table*}

    \section{Projected Wave-Functions and Variational Monte Carlo}
    
        The spin-liquid wave-functions are obtained from the following non-interacting mean-field Hamiltonian
        \begin{equation}
        {\cal H}_{{\rm MF}} =
        \sum_{(i,j),\alpha}t_{ij}c_{i,\alpha}^{\dagger}c_{j,\alpha} + 
        \sum_{(i,j)}\Delta_{ij}(c^{\dagger}_{i,\uparrow}c^{\dagger}_{j,\downarrow}+ {\rm H.c.})
        +\sum_{i}\Bigg\{\mu \sum_{\alpha}c_{i,\alpha}^{\dagger}c_{i,\alpha}
        +\zeta (c_{i,\uparrow}^{\dagger}c_{i,\downarrow}^{\dagger}+{\rm H.c.})\Bigg\},
        \label{eqn:MF-SL}
        \end{equation}
        where $t_{ij}=\langle c_{i,\alpha}^{\dagger}c_{j,\alpha} \rangle$ and 
        $\Delta_{ij}=\langle c_{i,\uparrow}c_{j,\downarrow}\rangle$.
        In order to construct $\mathbb{Z}_{2}$ QSLs in addition to $\mathrm{U}(1)$, we include singlet pairing ($\Delta_{ij}=\Delta_{ji}$) on links and on-site pairing $\zeta$, in addition to the hopping terms $t_{ij}$ and chemical potential $\mu$.
        The different symmetry-allowed patterns of the distribution of phases of $t_{ij}$, $\Delta_{ij}$, $\mu$ and $\zeta$ correspond to distinct spin liquids~\cite{sonnenschein:2024}.
        When a particle-hole transformation is performed on spin-down electrons only
        \begin{align}
        &c^{\dagger}_{i,\downarrow} \to c_{i,\downarrow},&
        &c^{\dagger}_{i,\uparrow} \to c^{\dagger}_{i,\uparrow},
        \end{align}
        the mean-field Hamiltonian~(\ref{eqn:MF-SL}) commutes with the total number of particles. Therefore the uncorrelated state is defined by filling suitable single-particle orbitals. Boundary conditions should be taken in order to have a unique state (i.e., filling all orbitals in a shell with the same mean-field energy). Periodic (P) and anti-periodic (A) boundary conditions along the {\bf a}$_1$ and {\bf a}$_2$ lattice vectors (see Fig.~\ref{fig:lattice}) can be considered, leading to four choices: [P,P], [P,A], [A,P], and [A,A] of boundary conditions. We employ [P,P] boundary conditions.
        
        However, $|\Psi_{\rm MF}\rangle$ lives in the enlarged (i.e., fermionic) Hilbert space and, in order to obtain a {\it bonafide} wave-function for spins, one must include fluctuations about the mean-field state. In this respect, an accurate treatment of all (spatial and temporal) fluctuations becomes important. On a lattice system, it proves impossible to analytically treat all these fluctuations in an accurate manner and one has to resort to approximate methods. Temporal fluctuations of the Lagrange multiplier $\mu$ are particularly important, since they enforce the one-fermion-per-site constraint. 
        
        Our variational wave functions for the spin model are thus defined as
        \begin{equation}\label{eqn:physical-wf}
        |\Psi_{\rm var}\rangle = \mathcal{P}_{G}|\Phi_0\rangle.
        \end{equation}
        Here, $|\Phi_0\rangle$ is an uncorrelated wave-function that is obtained as the ground state of a {\it generic} non-interacting Hamiltonian [Eq.~(\ref{eqn:MF-SL})], while $\mathcal{P}_{G}=\prod_{i}(n_{i,\uparrow}-n_{i,\downarrow})^2$ is the Gutzwiller projector.
        Note that $|\Phi_0\rangle$ is obtained without any self-consistent requirement, as in the mean-field approach, but is found by minimizing the energy in presence of the Gutzwiller projector.
        In this case, a Monte Carlo sampling is needed in order to compute any expectation values over variational states, since the resulting wave function includes strong correlations among the fermionic objects. 
        
        The variational parameters in the spin wave-function of Eq.~(\ref{eqn:physical-wf}) are optimized using an implementation of the stochastic reconfiguration (SR) optimization method~\cite{Sorella-2005,Yunoki-2006}. This allows us to obtain an extremely accurate determination of variational parameters. Indeed, small energy differences are effectively computed by using a correlated sampling, which makes it possible to strongly reduce statistical fluctuations.
        The close energetic competition between the two gapped $\mathbb{Z}_{2}$ QSLs at P1 and P2 highlights the indispensability of this approach.

\end{document}